\documentclass[]{interact}

\usepackage{epstopdf}
\usepackage{xcolor}
\usepackage[caption=false]{subfig}

\usepackage[numbers,sort&compress,merge]{natbib}

\bibpunct[, ]{[}{]}{,}{n}{,}{,}

\theoremstyle{plain}

\theoremstyle{definition}

\theoremstyle{remark}

 \usepackage{ragged2e}
\newcommand{\bp}{\beta p}
\newcommand\beq{\begin{equation}}
\newcommand\eeq{\end{equation}}
\newcommand\beqa{\begin{eqnarray}}
\newcommand\eeqa{\end{eqnarray}}
\newcommand{\nn}{\nonumber\\}
\def\bal#1\eal{\begin{align}#1\end{align}}
\newcommand{\dd}{\mathrm{d}}

\begin{document}


\title{On a conjecture concerning the Fisher--Widom line and the line of vanishing excess isothermal compressibility in simple fluids}

\author{
\name{Ana M. Montero\textsuperscript{a}\thanks{CONTACT Ana M. Montero. Email: anamontero@unex.es; \'Alvaro Rodr\'{\i}guez-Rivas. Email: arodriguezrivas@us.es. Web: https://arodriguez-rivas.weebly.com/; Santos B. Yuste. Email: santos@unex.es. Web: https://fisteor.cms.unex.es/investigadores/santos-bravo-yuste/; Andr\'es Santos. Email: andres@unex.es. Web: https://fisteor.cms.unex.es/investigadores/andres-santos/; Mariano L\'{o}pez de Haro. Email: malopez@unam.mx. Web: https://www.ier.unam.mx/academicos/mlh/}, \'Alvaro Rodr\'{\i}guez-Rivas\textsuperscript{b}, Santos B. Yuste\textsuperscript{a,c}, Andr\'es Santos\textsuperscript{a,c} and  Mariano L\'{o}pez de Haro\textsuperscript{a,d}}
\affil{\textsuperscript{a}Departamento de F\'isica, Universidad de Extremadura, 06006 Badajoz, Spain \\ \textsuperscript{b}
Departamento de Matem\'atica Aplicada II, Escuela T\'ecnica Superior de Ingenier\'ia, Universidad de Sevilla, Camino de los Descubrimientos s/n, 41092,  Seville, Spain
\\ \textsuperscript{c}Instituto de Computaci\'on Cient\'ifica Avanzada (ICCAEx), Universidad de Extremadura, 06006 Badajoz,
Spain \\
\textsuperscript{d}On sabbatical leave from Instituto de Energ\'{\i}as Renovables, Universidad Nacional Aut\'onoma de M\'exico (U.N.A.M.),
Temixco, Morelos 62580, M{e}xico.}
}

\maketitle

\begin{abstract}
In the statistical mechanics approach to liquid-state theory, understanding the role of the intermolecular potential in determining thermodynamic and structural properties is crucial. The Fisher--Widom (FW) line, which separates regions in the temperature vs density plane where the decay of the total correlation function is monotonic or oscillatory, provides insights into the dominance of the attractive or repulsive part of the interactions. Stopper et al. have recently conjectured [J. Chem. Phys. \textbf{151}, 014501 (2019)] that the line of vanishing excess isothermal compressibility approximates the FW line in simple fluids.
Here, we investigate this conjecture using the Jagla potential and also explore the line of vanishing excess pressure. We employ theoretical approximations and Monte Carlo simulations to study one-dimensional and three-dimensional systems. While exact results for the one-dimensional case do not support the conjecture, our Monte Carlo simulations for the three-dimensional fluid validate it.
Our findings not only contribute to the understanding of the relationship between the three transition lines but also provide valuable insights into the thermodynamic and structural behaviour of simple fluids.
\end{abstract}

\date{\today}




\justify

\section{Introduction}

In the statistical mechanics approach to the theory of liquids, a key goal is to be able to account for the bulk macroscopic properties of a given system in terms of the nature of the intermolecular interaction potential. In general, in order to capture the essential physics of real systems, models of such potential for simple fluids (which are taken to be spherically symmetric and pairwise additive) involve strong repulsion at short distances and weak attraction at  longer distances. Therefore, it is reasonable to try to assess the role played by the repulsive   and  attractive parts of the potential in determining the thermodynamic and structural properties of the fluid.
There is already a  fair amount of work in this direction reported in the literature \cite{FW69,MS90,EHHPS93,LEHH94,VRL95,BLE96,DE00,TCRV02,S02b,TCV03,XKBCPSS05,SD05,BBFRT13,SBHPG16,WSAE18,HRYS18,SHRE19,MS19,MFV22,LJ23,A23}.

Perhaps the simplest example of such an assessment in the case of the thermodynamic properties of fluids at low density is provided by the temperature-dependence of the second virial coefficient, $B_2(T)$. When the temperature is high enough and then the repulsive part of the intermolecular potential is dominant, $B_2(T)$ is positive and the pressure in the fluid is greater than that of an ideal gas. On the other hand, if the dominant part is the attractive one (at low enough temperatures), then $B_2(T)$ is negative and the pressure in the fluid is smaller than the one of an ideal gas. In fact, there is a particular value of the temperature, the Boyle temperature $T_B$, at which $B_2(T_B)=0$, implying that the pressure of the low-density fluid coincides with the one of the ideal gas and the repulsive and attractive interactions cancel each other out.

Another example related to the thermodynamic properties is the compressibility factor defined as $Z(\rho,T)={p}/{\rho k_B T}$, where $p$ is the pressure,  $\rho$ is the number density, $k_B$ is the Boltzmann constant and $T$ is the absolute temperature. As is well known, $Z=1$ for an ideal gas. When the attractive part of the potential dominates (low enough temperatures and/or densities), then $Z$  tends to be smaller than $1$, while if the repulsive part dominates (high enough temperatures and/or densities),  $Z$ tends to be greater than $1$. In the phase diagram of a simple fluid, the line in the temperature vs density plane separating the region where $Z<1$ from the one in which $Z>1$ is called the Zeno line \cite{BH90}. It is generally assumed to be an almost  straight line that starts at the Boyle temperature  and  ends by crossing the density axis at the so-called Boyle density  $\rho_B$, which is the value of the density obtained by extrapolating the coexistence curve into the low-temperature region beyond the triple point.
However, very recently Paterson et al.~\cite{PBL24} have found that, for both attractive square-well fluids with varying well-widths  and Mie $n$-6 fluids with different repulsive exponents $n$, irrespective of the values of the well-width or of the repulsive exponent, the corresponding Zeno lines are curved. We will come back to this point later on.

The value of another thermodynamic quantity, the isothermal susceptibility (or reduced isothermal compressibility) $\chi_T(\rho,T)=k_B T \left({\partial \rho}/{\partial p}\right)_T$, which is equal to $1$ for an ideal gas, also serves to indicate whether it is the attractive part of the potential the one that dominates (when $\chi_T > 1$) or whether the repulsive part is the dominant one (when  $\chi_T < 1$). The line in the phase diagram with $\chi_T=1$ (which also starts at the Boyle temperature in the temperature vs density plane) separates the regions where either part of the potential dominates from the perspective of the isothermal compressibility. The    line $\chi_T=1$ has been referred to in the literature as  the `line of vanishing excess isothermal compressibility' \cite{SHRE19}. However, in analogy with the reasoning \cite{BH90} that led to coin the term `Zeno' line ($Z=1$), from here onwards,  and for reasons to be explained below, we will abbreviate the nomenclature and refer to the line $\chi_T=1$ as the `Seno' line.

The above discussion has focussed on qualitative arguments related to (in principle) measurable thermodynamic quantities. We now turn specifically to structural properties. The statistical mechanics expression for the compressibility factor in $d$ dimensions, as obtained from the virial route, gives $Z$ in terms of the intermolecular potential $\phi(r)$ and the radial distribution function $g(r)$ as \cite{HM13,S16}
\begin{equation}
\label{virRoute}
Z=1-\frac{\rho}{2 d k_BT}\int \dd\mathbf{r}\, r \frac{\dd \phi(r)}{\dd r}g(r),
\end{equation}
where $r$ is the distance and $\dd\mathbf{r}$ the differential of volume in $d$ dimensions.
Also, the statistical mechanics expression for the isothermal susceptibility coming from the compressibility route reads
\begin{equation}
\label{chi_T}
\chi_T=1+\rho\int \dd\mathbf{r}\, h(r)=S(0),
\end{equation}
where $h(r)=g(r)-1$ is the total correlation function and $S(k)=1+\rho \int \dd\mathbf{r}\, e^{-i\mathbf{k}\cdot\mathbf{r}} h(r)$ is the structure factor. The idea behind the nomenclature `Seno' line follows from the equality $S(0)=1$ along that line.

The role played by the attractive and repulsive parts of the potential on the structural properties of simple fluids is best exemplified by the study (first carried out by Fisher and Widom \cite{FW69} for one-dimensional lattice-continuum models) of the asymptotic decay  of the total correlation function. In fact, the effect of a dominant repulsive part manifests itself in a damped oscillatory decay,  while the decay is monotonic if the dominant part is the attractive one.  The so-called Fisher--Widom  (FW) line in the temperature vs density plane of the phase diagram is the line that separates these two regions, namely the region in which the asymptotic decay of $h(r)$ is monotonic and the region in which it is damped oscillatory.

Although not directly linked to the dominance of the attractive or repulsive part of the potential, but rather to liquid-like behaviour in the supercritical region, there is another interesting line  in the temperature-density plane, the so-called Widom line \cite{XKBCPSS05}. This line is the locus of points of maximal response (for instance, maximal correlation length) for each temperature. As temperature decreases, the Widom line ends at the critical point, thus representing an extension of the coexistence line into the one-phase region.

The FW line has received a lot of attention and, recently, Stopper et al.\ \cite{SHRE19} have conjectured that  the Seno line should approximate well the FW line in simple fluids. They tested their hypothesis in a few models (square-well, hard-core Yukawa, sticky hard spheres and Asakura--Oosawa) and also located various lines relative to the gas-liquid  phase coexistence, as well as the Widom line.
It is the main aim of this paper to examine Stopper et al.'s conjecture by considering a particular model potential, the Jagla potential \cite{J99a} (hard core plus a linear repulsive ramp and a linear attractive ramp) given by
\begin{equation}
\label{phi(r)}
\phi(r)=\left\{
\begin{array}{ll}
\infty , & 0\leq r < \sigma, \\
\displaystyle{\frac{\epsilon_1 (\lambda_1 -r)-\epsilon_2 (r-\sigma)}{\lambda_1-\sigma}}, & \sigma<r\leq\lambda_1, \\
\displaystyle{-\frac{\epsilon_2 (\lambda_2 -r)}{\lambda_2-\lambda_1}}, & \lambda_1\leq r\leq\lambda_2, \\
0, & r \geq\lambda_2.
\end{array}
\right.
\end{equation}

This potential involves three lengths (the hard-core diameter $\sigma$ and the ranges $\lambda_1$ and $\lambda_2$) and two energies (the height $\epsilon_1$ of the repulsive ramp and the depth $\epsilon_2$ of the attractive well, both taken to be positive). Among its assets, it is able to predict multiple fluid transitions and some of the water-type thermodynamic and dynamic anomalies. Since the original work of Fisher and Widom \cite{FW69} was carried out for one-dimensional systems, while the conjecture was proposed for three-dimensional fluids \cite{SHRE19}, in this paper we will assess its value both for one-dimensional and three-dimensional Jagla fluids. Moreover, we will compare the FW and Seno lines with the Zeno line.
For further use, we introduce the dimensionless quantities
\beq
\rho^*=\rho\sigma^d,\quad T^*=\frac{1}{\beta^*}=\frac{k_BT}{\epsilon_2},\quad \epsilon_1^*=\frac{\epsilon_1}{\epsilon_2},
\eeq
as well as the characteristic distances
\beq
a_1= \frac{\lambda_1-\sigma}{\epsilon_1^*+1},\quad a_2=\lambda_2-\lambda_1.
\eeq
To illustrate our results for both the one-dimensional and the three-dimensional system, we  set
\beq
\label{param}
\frac{\lambda_1}{\sigma}=1.3,\quad \frac{\lambda_2}{\sigma}=1.6, \quad \epsilon_1^*=1.
\eeq
This choice for the values of the parameters is motivated by the fact that in the one-dimensional case the exact results require a nearest-neighbour interaction. On the other hand, for such values the three-dimensional Jagla fluid does not show a liquid-liquid phase separation \cite{GW06}.

This paper was prepared as an invited contribution to a special issue of Molecular Physics in honor of Luis F. Rull and Jos\'e Luis Fern\'andez Abascal. Apart from the fact that Luis addressed the problem of the location of the FW line for systems interacting through short-ranged potentials \cite{VRL95} and so our contribution is clearly aligned with the purpose of the special issue, we want to stress the personal connection of Luis with two of us (A.S. and A.R.R.). In this regard, we should mention that the first scientific paper that A.S. published \cite{BSR78} involved a collaboration with him. On the other hand, Luis was also the head of the group in which A.R.R. carried out his Ph.~D.\ thesis and together with Luis he published three papers \cite{RRRP13,RRRM14,RRR19},
which gave him the opportunity to start his career as a researcher in the statistical physics of liquids.

The paper is organised as follows. In Section~\ref{sec2}, we present the calculations pertaining to the one-dimensional Jagla fluid (in which case exact results may be derived) for the Zeno, Seno, FW and Widom lines. This is followed in Section~~\ref{sec3} by parallel calculations for the three-dimensional system, where we have used the theoretical rational-function approximation (RFA) \cite{SYH12,SYHBO13,S16} and Monte Carlo (MC) computer simulations. The paper is closed in Section~\ref{sec4} with a discussion of the results and some concluding remarks. Some mathematical details have been relegated to an Appendix.

\section{Test of the conjecture for the one-dimensional Jagla fluid. Exact results}
\label{sec2}
We begin with the case of the one-dimensional Jagla fluid. In order to evaluate the pertinence of the conjecture for this system, we will profit from the fact that the one-dimensional Jagla potential fulfills the requirements  that for one-dimensional fluids lead to explicit exact results for the thermodynamic and structural properties, namely that $\lim_{r \to 0}\phi(r)=\infty$, $\lim_{r \to \infty}\phi(r)=0$ and that each particle in the fluid interacts only with its two nearest neighbours if $\lambda_2\leq 2\sigma$. As exposed in Chapter 5 of Ref.~\cite{S16}, to which the reader is referred to for details, in these one-dimensional systems it is convenient to work with  the Laplace transforms of the radial distribution function $g(r)$ and of the Boltzmann factor $e^{-\beta \phi(r)}$ (where $\beta\equiv{1}/{k_BT}$),  namely
$G(s)=\int_0^\infty \dd r \, e^{-rs} g(r)$,  $\Omega(s,\beta)=\int_0^\infty \dd r\, e^{-rs} e^{-\beta \phi(r)}$. In fact, working in the isothermal-isobaric ensemble, one can express $G(s)$ in terms of $\Omega(s,\beta)$ as
\begin{equation}
\label{G(s)}
G(s)=\frac{\Omega'(\bp,\beta)}{\Omega(\bp,\beta)}\frac{\Omega(s+\bp,\beta)}{\Omega(s+\bp,\beta)-\Omega(\bp,\beta)},
\end{equation}
where $\Omega'(s,\beta)\equiv \partial_s\Omega(s,\beta)=-\int_0^\infty \dd r\, e^{-rs} r e^{-\beta \phi(r)}$. Furthermore,  the compressibility factor and the isothermal susceptibility may be expressed  as
\beq
\label{EOS_1D}
Z=-\bp \frac{\Omega'(\bp,\beta)}{\Omega(\bp,\beta)},\quad \chi_T=\frac{\Omega(\bp,\beta)\Omega''(\bp,\beta)}{[\Omega'(\bp,\beta)]^2}-1,
\eeq
where $\Omega''(s,\beta)\equiv \partial_s^2\Omega(s,\beta)=\int_0^\infty \dd r\, e^{-rs} r^2 e^{-\beta \phi(r)}$. Thus, in the   $\beta$ vs $\bp$ plane, the Zeno and Seno lines are given by the solutions to
\begin{subequations}
  \beq
  \label{Zeno1D}
\Omega(\bp,\beta)=-\bp \Omega'(\bp,\beta) \quad \text{(Zeno)},
\eeq
\beq
\label{Seno1D}
\Omega(\bp,\beta)\Omega''(\bp,\beta)=2[\Omega'(\bp,\beta)]^2 \quad \text{(Seno)}.
\eeq
\end{subequations}
The corresponding lines in the  $T$ vs $\rho$ plane are readily obtained from the equation of state $\rho=-\Omega(\bp,\beta)/\Omega'(\bp,\beta)=\bp$ (Zeno line) and $\rho=-\Omega(\bp,\beta)/\Omega'(\bp,\beta)=-2\Omega'(\bp,\beta)/\Omega''(\bp,\beta)$ (Seno line).

For the FW line, one needs the nonzero poles of $G(s)$, i.e. the roots of the equation $\Omega(s+\bp,\beta)=\Omega(\bp,\beta)$, with the least negative real part, since these will determine the asymptotic behaviour of the total correlation function $h(r)$. Near the FW line, the dominant poles are either a pair of complex conjugates ($s = -\zeta \pm i\omega$) or a real value ($s=-\kappa$), so that
\begin{equation}
\label{h(r)_1D}
h(r)\approx\left\{
\begin{array}{ll}
2|A_\zeta|e^{-\zeta r}\cos(\omega r+\delta), & \zeta<\kappa, \\
A_\kappa e^{-\kappa r}, & \zeta > \kappa, \\
\end{array}
\right.
\end{equation}
where $\delta$ is the argument of the residue $A_\zeta$ , i.e. $A_\zeta= |A_\zeta|e^{\pm i\delta}$ and $\kappa^{-1}$ is the correlation length. Once the poles have been computed, the FW line may readily be  obtained as the locus of points  where $\zeta=\kappa$, that is
\begin{subequations}
\label{FW_cond}
  \beq
  \label{FW_cond1}
  \text{Re}\left[\Omega(-\kappa\pm i\omega+\bp,\beta)\right]=\Omega(\bp,\beta),
  \eeq
  \beq
  \label{FW_cond2}
  \text{Im}\left[\Omega(-\kappa\pm i\omega+\bp,\beta)\right]=0,
  \eeq
  \beq
  \label{FW_cond3}
  \Omega(-\kappa+\bp,\beta)=\Omega(\bp,\beta).
  \eeq
\end{subequations}
Given a value of $\beta$, the solution to the set of Equations \eqref{FW_cond} yields the values of $\bp$, $\kappa$, and $\omega$ on the FW line. As before, the FW in the $T$ vs $\rho$ plane is obtained from  $\rho=-\Omega(\bp,\beta)/\Omega'(\bp,\beta)$.

As for the Widom line, it is obtained from Equation~\eqref{FW_cond3}, together with the condition $(\partial\kappa/\partial \bp)_{\beta}=0$. Deriving both sides of Equation~\eqref{FW_cond3} with respect to $\bp$, one can see that $(\partial\kappa/\partial \bp)_{\beta}=0$ yields
\beq
\label{Wline}
\Omega'(-\kappa+\bp,\beta)=\Omega'(\bp,\beta).
\eeq
The Widom line can be analytically continued as  a branch lying above the FW line by requiring that $\zeta^{-1}$ is maximal, i.e.  $(\partial\zeta/\partial \bp)_{\beta}=0$.

In the particular case of the Jagla potential, Equation~\eqref{phi(r)}, the function $\Omega(s,\beta)$ is
\beq
\label{omegaJagla}
\Omega(s,\beta)=-\frac{a_1 e^{-\beta^*\epsilon_1^*-\sigma s}}{\beta^*(1-a_1s/\beta^*)}
+\frac{s^{-1}e^{-\lambda_2 s}}{1+a_2s/\beta^*}
+\frac{(a_1+a_2) e^{\beta^*-\lambda_1 s}}{\beta^*(1-a_1s/\beta^*)(1+a_2 s/\beta^*)}.
\eeq

Up to this point, we now have all the necessary ingredients to compute the Zeno, Seno, FW and Widom lines for the one-dimensional Jagla fluid. But before doing that, and for the sake of completeness, we will take advantage of the simple form of the intermolecular potential $\phi(r)$ of this fluid, as given by Equation~(\ref{phi(r)}), to obtain explicitly its second virial coefficient. This will provide us with the means to compute also the Boyle temperature. The explicit analytic result for the second virial coefficient reads
\bal
B_2(T)=& -\int_0^\infty \dd r\left[e^{-\beta \phi(r)}-1\right]
=-\lim_{s\to 0}\partial_s\left[s\Omega(s,\beta)\right]\nn
=&\lambda_2-\frac{a_1(e^{\beta^*}-e^{-\beta^*\epsilon_1^*})+a_2(e^{\beta^*}-1)}{\beta^*}.
\label{secondvir}
\eal
For the choice given by Equation \eqref{param}, the Boyle temperature is $T_B^*\simeq 0.4758$.

\begin{figure}[t!]
    \centering
    \includegraphics[width=0.5\columnwidth]{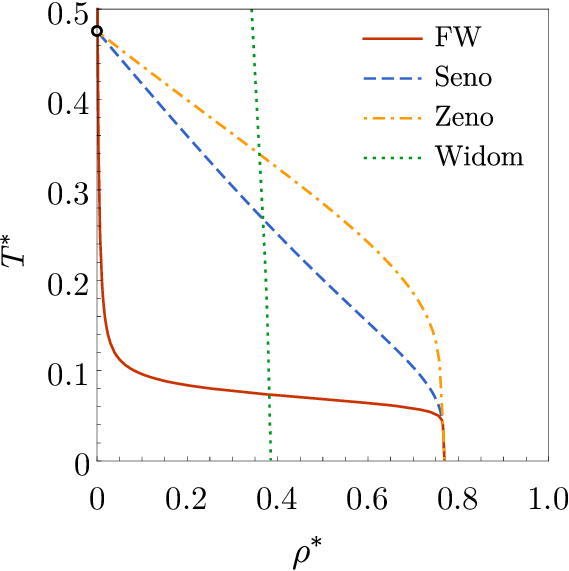}
       \caption{Zeno, Seno,  FW and Widom curves in the $T^*$ vs $\rho^*$ plane for a one-dimensional Jagla fluid with the parameters given in Equation~\eqref{param}. The open circle at $\rho^*=0$ represents the Boyle temperature $T_B^*\simeq 0.4758$.
    Below the Zeno line, one has $Z<1$, while $Z>1$ above it. Similarly, $\chi_T>1$ below the Seno line   and $\chi_T<1$ above it. Furthermore, below the FW line, the decay of $h(r)$ is monotonic, while it is oscillatory above it. The Widom line is the locus of points where the correlation length is maximal at a given temperature. While the Zeno, Seno and  FW lines terminate at the Boyle density $\rho_B^*=\sigma/\lambda_1\simeq 0.77$, the Widom line ends at $\rho_B^*/2\simeq 0.38$.
    }
    \label{fig1}
\end{figure}

In Figure \ref{fig1} we show the resulting Zeno, Seno, FW and Widom lines in the temperature vs density plane. Note that, while the Zeno and Seno lines do start at the Boyle temperature, the FW line diverges for $\rho\to 0$, despite the wrong impression one might get from the figure. Two more things are also worth pointing out at this stage. On the one hand, the Zeno line is not a straight line and ends at the Boyle density $\rho_B=\lambda_1^{-1}$; the same density is the zero-temperature end of the Seno and FW lines, while the Widom line terminates at $\rho_B/2$ (see the Appendix for a proof). On the other hand, it is clear that for this system the conjecture of Stopper et al.~\cite{SHRE19} concerning the Seno and FW lines is not sustained. Whether it will hold for the three-dimensional Jagla fluid  will be discussed in Section \ref{sec3}.

\section{The case of the three-dimensional Jagla fluid}
\label{sec3}

\subsection{Basics}

In this section we begin with the expression for the second virial coefficient of the three-dimensional Jagla fluid. This follows from the usual definition, namely
\bal
B_2(T)=&-\frac{1}{2}\int \dd\mathbf{r}\left[e^{-\beta \phi(r)}-1\right] \nn
=&\frac{2\pi}{3} \left\{\lambda_2^3-\frac{3a_2}{{\beta^*}^3}(e^{\beta^*}-1)\left[a_2^2+(a_2+\beta^*\lambda_1)^2\right]
+\frac{6a_2^2}{{\beta^*}^2}(a_2+\beta^*\lambda_1)+\frac{3a_2^3}{\beta^*}\right.\nn
&\left.
-\frac{3a_1}{\beta^*}\left(\lambda_1^2e^{\beta^*}-\sigma^2e^{-\beta^*\epsilon_1^*}\right)
+\frac{6a_1^2}{{\beta^*}^2}\left(\lambda_1e^{\beta^*}-\sigma e^{-\beta^*\epsilon_1^*}\right)
-\frac{6a_1^3}{{\beta^*}^3}\left(e^{\beta^*}-e^{-\beta^*\epsilon_1^*}\right)\right\}.
\label{B23d}
\eal
With the choice \eqref{param}, the Boyle temperature turns out to be $T_B^*\simeq 1.3879$.

The compressibility factor is obtained after substitution of Equation~\eqref{phi(r)}  into Equation \eqref{virRoute}. The result is
\beq
\label{Z3d}
Z=1+\frac{2\pi}{3}\rho\left[\sigma^3g(\sigma^+)+\frac{\beta^*}{a_1}\int_\sigma^{\lambda_1}\dd r \, r^3g(r)-\frac{\beta^*}{a_2}\int_{\lambda_1}^{\lambda_2} \dd r \, r^3g(r)\right],
\eeq
where $g(\sigma^+)$ is the contact value of the radial distribution function $g(r)$ of the three-dimensional Jagla fluid.
The isothermal susceptibility is still given by Equation \eqref{chi_T}, without any special simplification for the Jagla potential.

\subsection{Rational-function approximation}
In a previous paper \cite{HRYS18}, some of us presented a semi-analytical approach based on the RFA \cite{SYH12,SYHBO13,S16} to obtain $g(r)$, including its asymptotic behaviour for large $r$. The application of the RFA to the Jagla fluid was made by assuming that a \emph{discretised} version of the potential given in Equation \eqref{phi(r)} consisting in a hard core plus of a sequence of $n$ steps of heights $\varepsilon_j$ and widths $\sigma_j-\sigma_{j-1}$ (with the conventions $\sigma_0=\sigma$ and $\sigma_n=\lambda_2$), leads to essentially the same cavity function as the original Jagla potential. By considering the second virial coefficient and some representative cases, it was found that the choice $n=10$ proved to be a reasonable one, leading to  good agreement with MC simulation results. Such an agreement worsened as the density increased and/or the temperature decreased, especially near contact. But, even in those cases, the oscillations of $g(r)$ for larger distances were well accounted for, at least at a qualitative level.

The discretised version of the potential leads to the following result for the compressibility factor
\beq
\label{Zn}
 Z_n=1+\frac{2\pi}{3}\rho\sum_{j=0}^{n} \sigma_j^3 \Delta g(\sigma_j),
 \eeq
where $\Delta g(\sigma_j)=g(\sigma_j^+)-g(\sigma_j^-)$ is the jump of the radial distribution function at $r=\sigma_j$. For this jump, the RFA also provides an analytic expression which will be omitted here but may be found, together with the details of its derivation, in Ref.\ \cite{YSH22}. This serves to calculate the Zeno line. In the same reference, an analytic expression for the isothermal susceptibility $\chi_T$, which will again be omitted but will serve to calculate the Seno line, is also provided.

Now we turn to the asymptotic behaviour of the radial distribution function for large $r$, as obtained within the RFA approach. To that end, we take advantage of the fact that the RFA is formulated in Laplace space by expressing the Laplace transform $\mathcal{G}(s)=\int_0^\infty \dd r\, e^{-r s}r g(r)$ of $rg(r)$ as an explicit function of the Laplace variable $s$. Thus, in analogy with Equation \eqref{h(r)_1D}, we have
\begin{equation}
\label{asympt}
h(r)\approx \frac{1}{r}
\begin{cases}
2 |\mathcal{A}_\zeta| {e^{-\zeta r}}\cos(\omega r+\delta),& \zeta<\kappa,\\
\mathcal{A}_\kappa {e^{-\kappa r}},&\zeta>\kappa,
\end{cases}
\end{equation}
where either $s=-\zeta\pm i\omega$ or $s=-\kappa$ is the pole of $\mathcal{G}(s)$ with the least negative real part.

\subsection{Monte Carlo simulations}
We have conducted NVT MC simulations for the three-dimensional Jagla fluid with the parameters shown in Equation~\eqref{param}.
The number of particles has been fixed to $N = 10\,976$. To ascertain the Seno and FW lines,
$900$ independent simulations were performed for each considered density and temperature, starting from different initial physical states that were previously equilibrated. Each simulation consisted of $10^8$  MC steps, during which we measured the radial distribution function $g(r)$ every $20\,000$ steps with a spacing of $\Delta r=0.01\sigma$ up to a maximum distance of $r=12\sigma$. Finally, the results of $g(r)$ were averaged over all the simulations. For measuring the Zeno line, $200$ independent  simulations of $10^7$ MC steps each were conducted using a spacing of $\Delta r=0.001\sigma$. All simulations were carried out using a modified version of the DL\_MONTE software from the Collaborative Computational Project CCP5 \cite{PCP13,BGUSPPW19}, where the Jagla fluid potential was incorporated.

The density values utilised to determine the FW temperature were $\rho^*=0.20$, $0.25$, $0.30$, $0.35$ and $0.40$. Additionally, for the Seno line, we included $\rho^*=0.10$, and for the Zeno line, we incorporated $\rho^*=0.10$ and $\rho^*=0.50$. At each density, a varying number of temperature values were selected, typically with an interval of $\Delta T^*=0.05$.

Concerning the computation of the compressibility factor, we note from Equation (\ref{Z3d}) that it only requires knowledge of $g(r)$ in the interval from $r=\sigma$ to $r=\lambda_2$. Accurate values of $g(r)$ for a discrete set of points in this interval are relatively easy to get in the simulations and we used  the following discrete approximation
\beq
\label{Z3dm}
Z\approx 1+\frac{2\pi}{3}\rho\left[\sigma^3g(\sigma^+)+\frac{\beta^*}{a_1}\Delta r\sum_{\sigma\leq r_i\leq \lambda_1}r_i^3g(r_i)
-\frac{\beta^*}{a_2}\Delta r\sum_{\lambda_1\leq r_i\leq \lambda_2}r_i^3g(r_i)\right].
\eeq
From the numerical values of $Z$ at a given density, the associated Zeno temperature was obtained by interpolation to $Z=1$.

\begin{figure}[t!]
    \centering
    \includegraphics[width=\columnwidth]{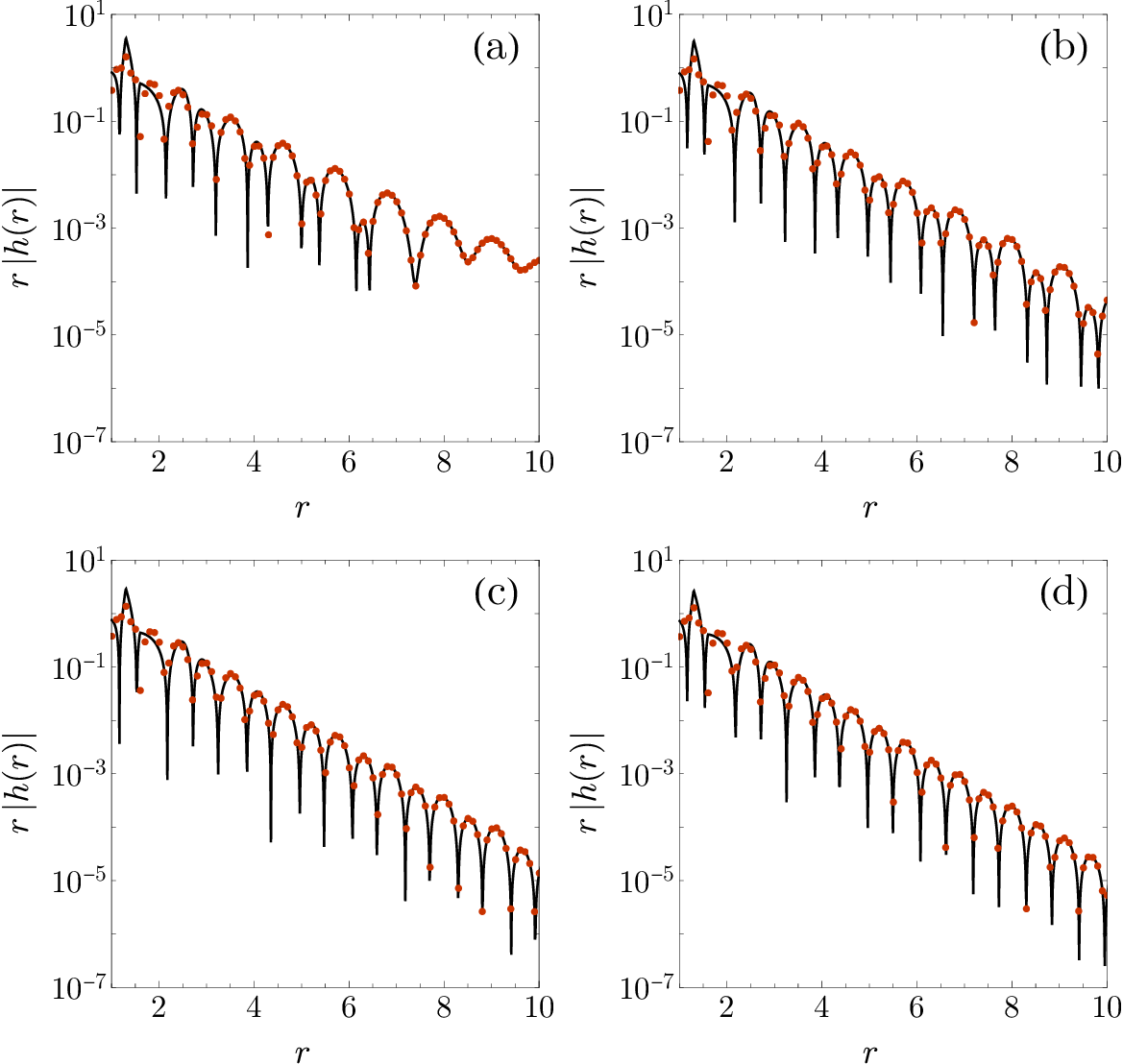}
    \caption{Plot of $r |h(r)|$ (in logarithmic scale), as predicted by the RFA,  for a three-dimensional Jagla fluid with the parameters given in Equation~\eqref{param} and $\rho^*=0.30$. The temperatures are (a) $T^*=0.60$, (b) $T^*=0.65$, (c) $T^*=0.70$  and (d) $T^*=0.75$. The solid lines correspond to the full approximation, while the circles have been obtained using the two leading poles. Note that the hard-core diameter $\sigma=1$ has been taken as the unit of length.}
    \label{fig2}
\end{figure}

The MC computation of $\chi_T=S(0)$ is a little bit more involved since the values of $g(r)$ for all $r$ are needed [cf.~Equation~\eqref{chi_T}]. What we have done is the following. The MC data for $g(r)$ between the distances $r=R_1$ and $r=R_2$ have been fitted to the functional form
\beq
\label{mC1}
g_{\text{asympt}}(r)\equiv g_\infty +A_\kappa \frac{e^{-\kappa r}}{r}+2|A_\zeta| \frac{e^{-\zeta r}}{r}\cos(\omega r+\delta).
\eeq
This form is based on the expected competition between the real and complex poles, as given by Equation \eqref{asympt}. Moreover, it must be pointed out that, due to unavoidable finite-size effects, the asymptotic value of $g(r )$ in the MC simulations does not necessarily tend to $1$, but rather to a value which we refer to as $g_\infty$, with $|g_\infty-1|\sim 10^{-4}$--$10^{-5}$.
With such an approximation, we then have evaluated $\chi_T$ as follows
\beq
\label{mC2}
\chi_T\approx 1+4\pi\rho \left[-\frac{\sigma^3}{3}+\Delta r\sum_{\sigma\leq r_i\leq R_2}r_i^2 h(r_i)+\int_{R_2}^\infty \dd r\, r^2h_{\text{asympt}}(r)\right],
\eeq
where now the MC values of the total correlation function are defined as $h(r)=g(r)-g_\infty$ and $h_{\text{asympt}}(r)=g_{\text{asympt}}(r)-g_\infty$.
Note that the integral $\int_{R_2}^\infty \dd r\, r^2h_{\text{asympt}}(r)$ may be obtained analytically, although we omit here its explicit expression.
We have checked that an optimal choice is $R_1=4\sigma$ and $R_2=7\sigma$.
Once we obtain $\chi_T$ for several temperatures at a given density, the Seno temperature is obtained by interpolation to $\chi_T=1$.

\begin{figure}[t!]
    \centering
    \includegraphics[width=\columnwidth]{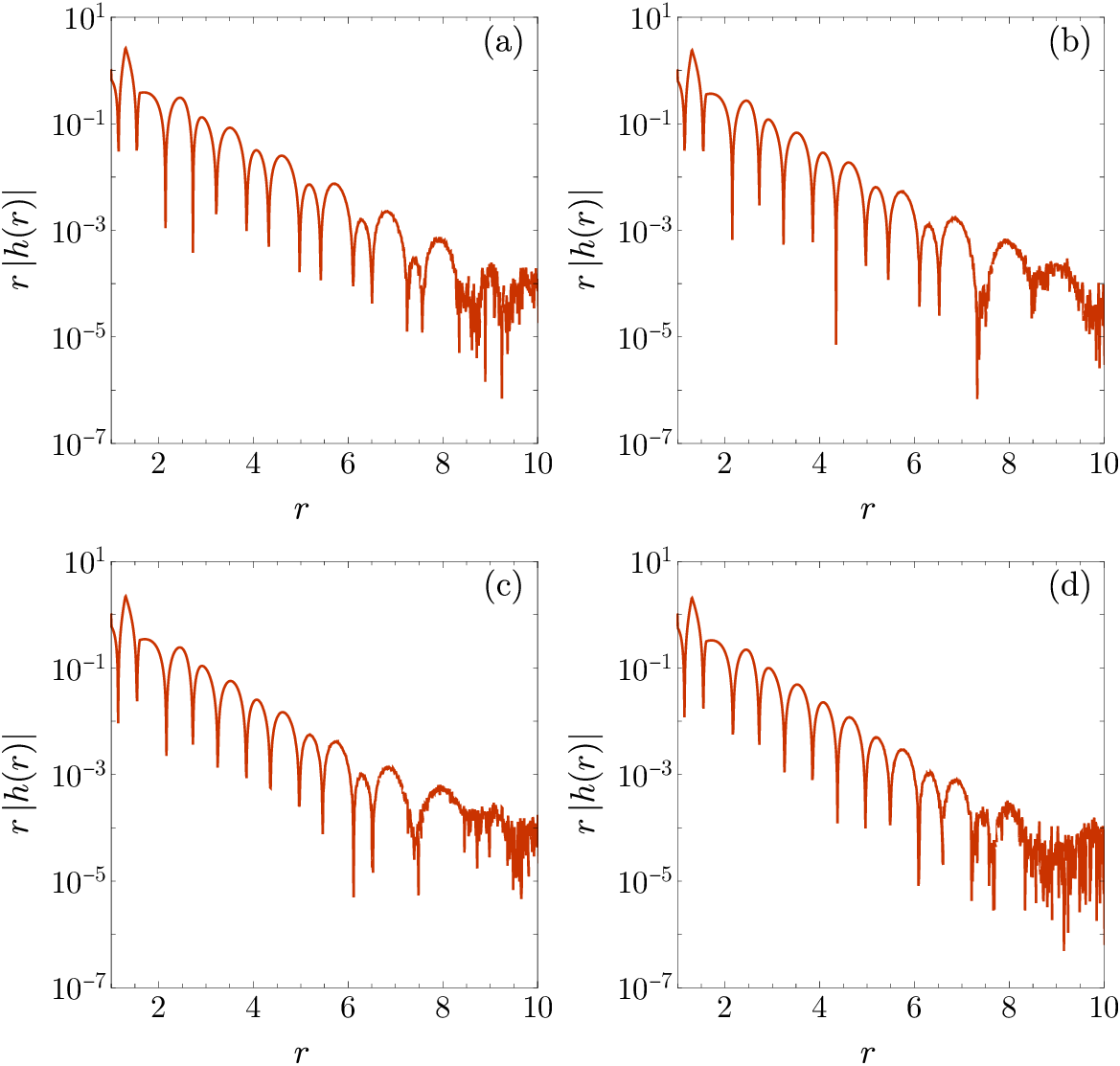}
    \caption{Plot of $r |h(r)|$ (in logarithmic scale), as obtained from our MC simulations,  for a three-dimensional Jagla fluid with the parameters given in Equation~\eqref{param} and $\rho^*=0.30$. The temperatures are (a) $T^*=0.75$, (b) $T^*=0.80$, (c) $T^*=0.85$  and (d) $T^*=0.90$.  Note that the hard-core diameter $\sigma=1$ has been taken as the unit of length.}
    \label{fig3}
\end{figure}

For the FW line, the main problem is how to know from the MC data of $g(r)$ at a given state $(\rho^*,T^*)$ sufficiently close to the line whether that state is above the line (region of oscillatory decay) or below it (region of monotonic decay).
If $g(r)$ were known with a good signal-to-noise ratio in the asymptotic large-$r$ domain, it would be in principle possible to assess whether the decay is oscillatory or monotonic since one of the two competing behaviours in Equation~\eqref{asympt} would dominate. However, the closer the state is to the line, the closer the values of $\kappa$ and $\zeta$ become. Consequently, larger distances are required to observe the prevalence of one of the two competing behaviours. In addition, it is worth noting that the amplitude $A_\kappa$ of the monotonic behaviour is typically smaller than the amplitude $2|A_\zeta|$ of the oscillatory behaviour. As a result, the oscillatory behaviour can overshadow the monotonic behaviour for intermediate distances, even if $\kappa<\zeta$, a feature that was also observed and reported by Stopper et al.~\cite{SHRE20} for patchy particles.
We have also noted that the fitting in Equation~\eqref{mC1}, although suitable for measuring $\chi_T$, lacks robustness in determining whether $\kappa<\zeta$ or $\kappa>\zeta$.

To establish a practical criterion that would provide us with at least a lower bound on the position of the FW line, we have turned to the RFA as a guide. As will be seen, this allows us to identify a signature of the monotonic-to-oscillatory transition in the behaviour of $r |h(r)|$ for distances smaller than, say, $r=8\sigma$.

Figure~\ref{fig2} shows $r |h(r)|$ (in logarithmic scale), as obtained from the RFA, for a density $\rho^*=0.30$ and  four temperatures: $T^*=0.60$, $0.65$, $0.70$  and $0.75$. For this density, the RFA temperature corresponding to the FW line is known to be $T^*=0.7315$. The first thing to note in this case is that the approximation with the two leading poles is able to capture the whole total correlation function for distances beyond $r\simeq 3\sigma$ for all temperatures.
Next, we note that the signature that one is sufficiently below the temperature corresponding to the FW line is that  `anomalous' neighbouring nodes appear [cf.\ Figures~\ref{fig2}(a--c)]. These nodes, which eventually disappear for large enough distances [although oscillations may still be seen, cf.\ Figure~\ref{fig2}(a)],   exhibit an anomalous behavior: their separation is smaller than that of neighbouring nodes and their maxima always fall below that of the neighbouring peaks [cf.\ Figure~\ref{fig2}(a)].  As the temperature increases, remaining below the FW line, the anomalous nodes become progressively less apparent within the range  $r<10\sigma$ [cf.\ Figures~\ref{fig2}(b,c)]. Finally, when one is close to or above the temperature corresponding to the FW line, the nodes become regular [cf.\ Figure~\ref{fig2}(d)]. According to our criterion, one would conclude that $T^*=0.65$, or even $T^*=0.70$,  are lower-bound estimates for the temperature of the FW line when $\rho^*=0.30$, which agrees with the true FW temperature $T^*=0.7315$ predicted by the RFA for $\rho^*=0.30$.

We have applied the criterion above to obtain (lower-bound) estimates of the FW temperatures from our MC values of $g(r)$. As an illustration, Figure~\ref{fig3} shows the MC values of $r |h(r)|$ for a density $\rho^*=0.30$ and  the temperatures $T^*=0.75$, $0.80$, $0.85$  and $0.90$.
We have estimated the right values of $g_\infty$ by requiring that the fluctuations of $r|h(r)|$ in the region $r>R_2=7\sigma$ are maximised and so what one is seeing at such distances is the statistical error associated with the numerical data and not the effect of the value of $g_\infty$.
Following the above rationale, we have determined the value of $g_\infty$ for all the results of our simulations. For instance, at $\rho^*=0.30$
we find $g_\infty=0.99995$, $1$, $1.00005$ and $1.00005$ for $T^*=0.75$, $0.80$, $0.85$  and $0.90$, respectively. Combining the inclusion of $g_\infty$ and the previous criterion, we find that $T^*=0.85$ is a lower-bound estimate of the FW temperature from the simulation data for $\rho^*=0.30$.

\begin{figure}[t!]
    \centering
    \includegraphics[width=0.5\columnwidth]{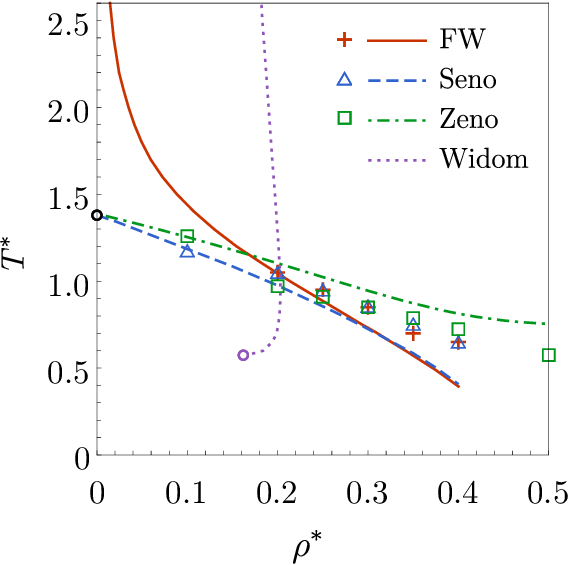}
    \caption{Zeno, Seno,  FW and Widom curves in the $T^*$ vs $\rho^*$ plane for a three-dimensional Jagla fluid with the parameters given in Equation~\eqref{param}. The lines are RFA predictions and the symbols represent estimates obtained from our MC simulations. The open circle at $\rho^*=0$ represents the Boyle temperature $T_B^*\simeq 1.3879$.
    Below the Zeno line, one has $Z<1$, while $Z>1$ above it. Similarly, $\chi_T>1$ below the Seno line   and $\chi_T<1$ above it. Furthermore, below the FW line, the decay of $h(r)$ is monotonic, while it is oscillatory above it. The Widom line is the locus of points where the correlation length is maximal at a given temperature. }
    \label{fig4}
\end{figure}

\subsection{Results}
In order to set the proper perspective for the assessment of our findings, in Figure \ref{fig4} we show the resulting Zeno, Seno and FW lines for the three-dimensional Jagla fluid, as obtained both from the RFA approach (with a discretisation of $n=10$ steps) and from simulation. The Widom line predicted by the RFA \cite{HRYS18} is also included. It terminates at the critical point $(\rho_c^*,T_c^*)=(0.162,0.574)$, which slightly shifts to $(\rho_c^*,T_c^*)=(0.160,0.577)$ if $n=20$ is employed.

One immediately notices two things. On the one hand, at least for $\rho^*= 0.20$, $0.25$, $0.30$, $0.35$ and $0.40$, the overlap in the simulation data indicates that the conjecture of Stopper et al. \cite{SHRE19} is fulfilled reasonably well in this density range. Moreover and remarkably, although to a lesser extent, there is also reasonable quantitative agreement between the simulation data points of the FW line and those of the Zeno line. While the RFA approach captures qualitatively the proximity of the FW and Seno lines for $\rho^*\geq 0.20$, it fails to do so in the case of the FW and the Zeno lines. In fact, the results of the RFA approach always overestimate the values of the points of the Zeno line for that density range. On the other hand, it is clear that, as expected, quantitatively the performance of the RFA approach worsens for the higher densities and the lower temperatures. In fact, as reflected in Figure \ref{fig4}, beyond $\rho^*\simeq 0.40$ the RFA numerical calculations are not reliable for both the Seno and the FW lines and hence they have not been included.

\begin{figure}[t!]
    \centering
    \includegraphics[width=\columnwidth]{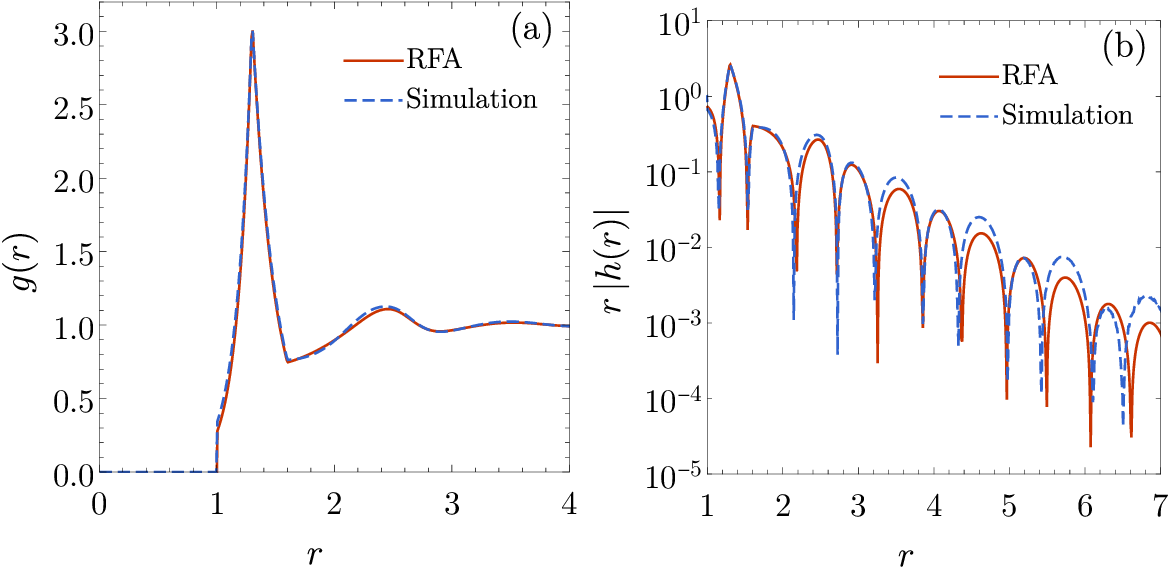}
       \caption{Plot of (a) $g(r)$  and (b) $r|h(r)|$ (in logarithmic scale) for a three-dimensional Jagla fluid with the parameters given in Equation~\eqref{param} and $\rho^*=0.30$,  $T^*=0.75$. The solid lines correspond to the RFA, while the dashed lines represent MC simulation data. Note that the hard-core diameter $\sigma=1$ has been taken as the unit of length.}
    \label{fig5}
\end{figure}

To illustrate how the discrepancies between the RFA and MC simulations for the transition lines are consistent with a reasonable global agreement in the radial distribution function,  we compare the RFA and MC values of $g(r)$ and $r|h(r)|$ at a density $\rho^*=0.30$ and a temperature $T^*=0.75$ in Figure~\ref{fig5}. These conditions correspond to the scenarios depicted in Figures~\ref{fig2}(d) and \ref{fig3}(a), respectively. A remarkable overall agreement is observed, although the RFA tends to slightly underestimate $g(r)$ within the interval $\sigma\leq r\leq \lambda_1$ and near the second maximum. Considering Equations~\eqref{chi_T} and \eqref{Z3d}, this suggests that the RFA tends to underestimate the values of $Z$ and $\chi_T$. Consequently, this leads to an upward shift of the Zeno line and a downward shift of the Seno line with respect to the MC values. It is also evident from Figure~\ref{fig5}(a) that $g(r)\simeq 1$ for $r>4\sigma$ if $\rho^*=0.30$ and $T^*=0.75$. This makes it rather challenging to determine whether the asymptotic decay is monotonic or oscillatory. In the case of the RFA, we know from the pole analysis of the Laplace transform $\mathcal{G}(s)$ that the decay is oscillatory, while our criterion suggests that the decay of the MC data is monotonic. This distinction is clearly apparent in Figure~\ref{fig5}(b).

The previous observations indicate that, in view of its already known limitations, the good qualitative (and even quantitative) performance of the RFA approach observed in a certain region of the phase diagram may be lost under very stringent conditions of high density and low temperature.

\section{Discussion}
\label{sec4}

In this paper we have addressed one aspect of the role played by the attractive and repulsive parts of the intermolecular potential on the thermodynamic and structural properties of fluids. In particular, we have dealt with a conjecture, introduced by Stopper et al.\ \cite{SHRE19}, concerning the proximity of the FW line and the line of vanishing excess isothermal compressibility (for which we have coined the name Seno line) in simple fluids. To test the validity of such a conjecture, we have taken the intermolecular potential to be the Jagla potential \cite{J99a}, since this model potential may account for multiple fluid transitions and for some of the thermodynamic and dynamic anomalies observed in water. Both the one-dimensional  and the three-dimensional fluids have been considered. The second virial coefficient and the Zeno line, which also reflect the role played by the attractive and repulsive parts of the potential, have been obtained for these model fluids too. For the sake of illustration, we have taken in the two systems the set of parameters displayed in Equation~\eqref{param}.

The consideration of the one-dimensional system allowed us to derive exact results for all four lines. In this instance, we find that the conjecture is not satisfied (cf.\ Figure~\ref{fig1}). Since the Seno line is defined by the condition $\int_0^\infty \dd r\, r^n h(r)=0$, with $n=0$ for one-dimensional systems, one might reasonably wonder whether a modified condition with $n>0$ would emphasise the attractive part of the interaction and could serve as a better proxy for the FW line. However, our findings (not shown) indicate that $n=1$ and $n=2$ produce just the opposite effect.

In the case of the three-dimensional system, we have obtained approximate theoretical results with the RFA approach and we have also carried out MC simulations. Our findings indicate that, in contrast to what we found for the one-dimensional Jagla fluid, the conjecture of Ref.~\cite{SHRE19} is satisfied reasonably well, at least for $\rho^*= 0.20$, $0.25$, $0.30$, $0.35$ and  $0.4$. This is not very surprising, since the criterion  of using the ideal-gas-like isothermal compressibility to estimate the FW line \cite{SHRE19} is actually a mean-field idea that should work best  when higher dimensions are considered.
Interestingly, in the same density range we also find a proximity between the Zeno, Seno and FW lines. Whether this feature will hold also for other fluids is worth investigating. On the other hand, we also find that, while the RFA approach agrees qualitatively in the description of the density behaviour of the FW, Seno and Zeno lines, it overestimates in general the points on the Zeno line and fails to capture the proximity of the Zeno line with the other two lines in the density interval mentioned above. Furthermore, our analysis confirms that, although the RFA approach provides generally good results for the structural and thermodynamic quantities, it exhibits poor performance in accurately predicting the behaviour of the three transition lines, especially under conditions of high density and/or low temperature.

Finally, it is worth noting that the findings presented in this paper offer additional evidence of the impact of dimensionality (or confinement) on the thermodynamic and structural properties of fluids.

\section*{Acknowledgments}
The authors are grateful to the computing facilities of the Instituto de Computaci\'on Cient\'ifica Avanzada of the University of Extremadura (ICCAEx) and of C3UPO (University Pablo de Olavide), where the simulations were run.
M.L.H. thanks Universidad de Extremadura for its hospitality during his sabbatical year.

\section*{Disclosure statement}
No potential conflict of interest was reported by the authors.

\section*{Funding}

A.M.M., S.B.Y., A.S. and M.L.H. acknowledge financial support from Grant No.~PID2020-112936GB-I00 funded by MCIN/AEI/10.13039/501100011033, and from Grant No.~IB20079  funded by Junta de Extremadura (Spain) and by ERDF ``A way of making Europe.''
A.M.M. is also grateful to the Spanish Ministerio de Ciencia e Innovaci\'on for a predoctoral fellowship PRE2021-097702.
A.R.R. acknowledges financial support from Grant No.~PID2021-126348NB-I00 funded by MCIN/AEI/10.13039/501100011033.


\bibliographystyle{tfo}

\appendix

\section{Low-temperature limit of the Zeno, Seno, FW and Widom lines for the one-dimensional fluid}
In this Appendix we consider the one-dimensional Jagla fluid and analyze the limit $\beta^*\to\infty$ of the Zeno, Seno, FW and Widom lines, proving that the first three of them end at the Boyle density $\rho_B=\lambda_1^{-1}$, while the Widom line ends at $\rho_B/2$.

\subsection{Zeno line}
If $\beta^*\to\infty$ but $s\sim 1$, from Equation~\eqref{omegaJagla} we have
\beq
\Omega(s)\to a\frac{e^{\beta^*-\lambda_1 s}}{\beta^*},\quad \Omega'(s)\to - a\frac{e^{\beta^*-\lambda_1 s}}{\beta^*}\lambda_1,
\eeq
where $a\equiv a_1+a_2$ and, for simplicity,  we have omitted the argument $\beta$ in $\Omega(s,\beta)$.
Thus, Equation~\eqref{Zeno1D} yields $\bp\to \lambda_1^{-1}$ for the Zeno line. Since $Z=1$ on that line, we have $\rho\to \lambda_1^{-1}$.

\subsection{Seno line}
Now we are interested in the region where $\beta^*\to \infty$ and $s\to 0$ with $s^3\sim \beta^*e^{-\beta^*}$. Under those conditions,
Equation~\eqref{omegaJagla} becomes
\beq
\label{A2}
\Omega(s)\to s^{-1}+a\frac{e^{\beta^*}}{\beta^*}\left[1-\lambda_1 s+(a_1-a_2)\frac{s}{\beta^*}
+\frac{1}{2}\lambda_1^2s^2+\cdots\right].
\eeq
Therefore,
\beq
\Omega(s)\to a\frac{e^{\beta^*}}{\beta^*},\quad \Omega'(s)\to -a\lambda_1\frac{e^{\beta^*}}{\beta^*},\quad \Omega''(s)\to 2s^{-3}+a\lambda_1^2\frac{e^{\beta^*}}{\beta^*}.
\eeq
From Equation~\eqref{Seno1D} we get
\beq
\bp\to \left(\frac{2\beta^*}{a\lambda_1^2}\right)^{1/3}e^{-\beta^*/3}.
\eeq
Finally, Equation~\eqref{EOS_1D} gives $Z\to \lambda_1\bp$, i.e. $\rho\to\lambda_1^{-1}$.

\subsection{FW line}
In this case, we have to deal with Equations~\eqref{FW_cond}. Taking the limit $\beta^*\to\infty$, one can see that $\bp\to 0$ and $\kappa-\bp\to 0$. Then, taking into account Equation~\eqref{A2}, Equation~\eqref{FW_cond3} yields
\beq
\frac{1}{\bp}+\frac{1}{\kappa-\bp}=a\lambda_1 \kappa\frac{e^{\beta^*}}{\beta^*},
\eeq
which implies
\beq
\label{A6}
\kappa-\bp\to  \frac{ \beta^*e^{-\beta^*}}{a\lambda_1\bp}.
\eeq
Analogously, from Equation~\eqref{FW_cond2} one gets
\beq
\label{A7}
\omega\to \frac{2\pi}{\lambda_1}\left(1+\frac{a_1-a_2}{\lambda_1 \beta^*}\right).
\eeq
Finally, $\bp$ is determined by inserting Equations~\eqref{A6} and \eqref{A7} into Equation~\eqref{FW_cond1} and taking the limit $\beta^*\to\infty$. After some algebra, the result is
\beq
\label{23a}
\bp \to 2\pi^2\frac{a_1^2+a_2^2}{\lambda_1^3} \beta^{*-2}.
\eeq
Again, from Equation~\eqref{EOS_1D} we have $Z\to \lambda_1\bp$, implying $\rho\to\lambda_1^{-1}$.

\subsection{Widom line}
In the low-temperature regime, the Widom line necessarily resides below the FW line, thereby rendering the damping coefficient $\kappa$ determined by Equation~\eqref{FW_cond3}. For a fixed value of $\bp$, it can be seen that $\kappa-\bp\sim \beta^* e^{-\beta^*}$ as $\beta^*\to\infty$. Consequently, by substituting $\Omega(-\kappa+\bp)\to -(\kappa-\bp)^{-1}+a e^{\beta^*}/\beta^*$ and $\Omega(\bp)\to a e^{-\bp\lambda_1}e^{\beta^*}/\beta^*$ into Equation~\eqref{FW_cond3}, we obtain
\beq
\kappa\to\bp+\frac{\beta^* e^{-\beta^*}}{a}\left(1-e^{-\bp\lambda_1}\right)^{-1}.
\eeq
Now, the Widom condition $(\partial\kappa/\partial\bp)_\beta=0$ yields
\beq
\bp\to \sqrt{\frac{\beta^*}{a\lambda_1}}e^{-\beta^*/2}.
\eeq
Therefore, $\Omega(\bp)\to a{e^{\beta^*}}/{\beta^*}$, $\Omega'(\bp)\to -2a\lambda_1{e^{\beta^*}}/{\beta^*}$, which implies $\rho\to (2\lambda_1)^{-1}$, i.e. half the Boyle density. This result is analogous to the one previously obtained for the one-dimensional triangle-well fluid \cite{MS19}.

\end{document}